# The Policy Paradox: Government Debt Servicing and Local Bank Risk Growth

**LI Yan**

**1155191643@link.cuhk.edu.hk**

**The Chinese University of Hong Kong**

**Abstract:**

The issue of local government debt is widely recognized as one of the "gray rhinos" affecting the stable development of China's economy. Government debt can transmit risks to local banks, which are among the primary holders of local debt, thereby triggering systemic financial risks. Consequently, exploring debt resolution pathways and evaluating the systematic effects of debt servicing policies has become critically important. This study employs panel data from 348 local commercial banks across 29 provincial-level administrative regions in China from 2010 to 2023, and constructs a difference-in-differences (DID) model to investigate the impact of the State Council's special supervision of debt servicing on local bank risks. The findings indicate that the government's debt servicing policy essentially represents a shift of government debt from explicit to implicit forms, significantly increasing the risks faced by local banks and producing outcomes contrary to the policy's original intent. This effect is particularly pronounced for rural commercial banks and banks with high customer concentration and fewer branches. Mechanism analysis reveals two key insights. First, local banks are heavily influenced by local government control; the government's debt servicing requires banks to support the government by purchasing government bonds and other financial instruments, which leads to a deterioration in asset quality and an expansion of risk exposure. Second, government debt crowds out private credit from local banks, weakening the region's repayment capacity and ultimately increasing bank risk. Our research uncovers the counterintuitive effects of government debt servicing and offers corresponding policy recommendations.

## 1. Introduction

The long-term growth of China's economy has heavily relied on government-led investments. Since the reform and opening-up, local governments have raised development funds through debt financing to drive economic construction (Li & Zhou, 2005). The 1994 tax-sharing reform in China significantly curtailed the fiscal resources of local governments while delegating more responsibilities to them. This contradiction resulted in long-term funding shortages for local governments, compelling them to borrow heavily due to immense fiscal pressures (Zhang, 2013). Meanwhile, as local officials are primarily appointed by higher-level authorities and their performance evaluations are heavily tied to local economic performance, they are motivated to accelerate economic growth (Jin et al., 2005). Consequently, local government debt has expanded rapidly.

In the process of debt expansion, Chinese local governments established various financing platforms to meet their funding needs, with local commercial banks (LCBs) being a crucial component (Gao et al., 2021). Since 1995, these banks, including urban commercial banks and rural commercial banks, have been established across the country. These banks typically have smaller asset sizes, operate locally, and are often partially or wholly owned by local governments. Regional branches of large and state-owned banks do not fall into this category. Local governments, as major shareholders, exert significant control over these banks. Additionally, even private or government-minority-owned banks are subject to considerable government intervention due to the dominant influence of officials in local economic development (Qian et al., 2015).

LCBs play a vital role in supporting government debt. On the one hand, local governments raise funds through instruments such as municipal bonds and trust products, with LCBs being key investors (Zhao et al., 2018). On the other hand, these banks provide direct loans to governments and their financing vehicle (LGFVs) (Gao et al., 2021). According to the China Government Debt Center (2025), by the end of 2024, commercial banks held approximately 75% of local government debt.

Excessive debt beyond repayment capacity can lead to defaults, potentially triggering contagion effects that impact other financial institutions and market participants (Zhu et al., 2024). As primary debt holders, LCBs' operational stability is closely tied to the size of government debt and related policies. Additionally, LCBs, acting as financial intermediaries between the local financial system and the real economy, play a critical role in fostering economic growth. Debt servicing policies that require banks to finance the government often increase off-balance-sheet implicit debt, transmitting debt risks to the real economy and jeopardizing financial stability.

To address these challenges, the central government introduced several policies. In July 2016, the State Council issued the "Notice on Further Improving Work Related to Private Investment," which mandated the Ministry of Finance to coordinate efforts to resolve government payments. A special supervision team was dispatched to seven provinces and cities, including Beijing, Liaoning, Anhui, Shandong, Henan, Hubei, and Qinghai, to oversee the implementation of these policies, leading to significant progress in debt servicing. In the meanwhile, provinces that are not under supervision have maintained a more stagnant debt servicing effort.

This study uses the issuance of this notice and the related supervisory actions as an exogenous shock to examine the impact of government debt servicing policies on LCB risk. Employing panel data from 349 LCBs in China from 2010 to 2023 and using a difference-in-differences (DID) model with year, individual, and regional fixed effects. We find that debt servicing policies significantly increase non-performing loan (NPL) ratios, particularly for rural commercial banks and banks with high customer concentration and limited branch networks.

Robustness tests confirm the persistence of these findings across different economic conditions, less debt-burdened regions, and alternative dependent variables.

Mechanism analysis indicates that debt servicing policy increases bank risks through two primary channels: (1) by turning banks into financing tools for the government, deteriorating asset quality; and (2) by crowding out private credit, reducing regional repayment capacity. Both pathways contribute to heightened bank risk.

This study makes three primary contributions to the literature. First, while previous research predominantly focuses on the effects of debt expansion, this study provides novel insights from the perspective of debt servicing. Second, it incorporates the risks taken by local banks into the analytical framework, offering a more comprehensive understanding of the dynamic relationship between debt and risk. Third, by designing new control experiments within supervised regions, the study mitigates endogeneity concerns associated with policy self-selection problems, thus enhancing result robustness. Finally, we elucidate the mechanisms through which debt servicing policies affect bank risk, contributing practical insights for policy formulation.

The remainder of the paper is structured as follows: Section 2 reviews the theoretical framework and related literature; Section 3 outlines the research design, including data selection, variable definitions, and model construction; Section 4 presents empirical findings, including robustness checks and heterogeneity analysis; Section 5 delves into the underlying mechanisms; and Section 6 concludes with policy recommendations and limitations.

## 2. Theoretical Analysis and Literature review

The 1994 tax-sharing reform significantly reduced the fiscal authority of local governments, leaving them with insufficient funds to sustain economic operations (Zhang, 2013). Meanwhile, the emphasis on economic performance as a key criterion for officials' evaluations increased local governments' motivation to borrow and drive economic growth (Jin et al., 2005). Under these two mechanisms, local governments have long needed external financing to fill fiscal gaps. However, from 1994 to 2015, China's Budget Law prohibited local governments from incurring any debt. Even though the 2015 amendment granted governments limited borrowing rights, the debt scale remained strictly capped (Qu et al., 2023). Against this backdrop, local governments turned to off-balance-sheet financing methods, leading to a sharp increase in implicit debt.

The Chinese government primarily borrows through the establishment of off-balance-sheet financing platforms known as Local Government Financing Vehicles (LGFVs). These LGFVs help governments circumvent legal restrictions and meet their financing needs (Gao et al., 2021), while also transferring fiscal risks into financial risks. LCBs play a crucial role in this process because they act as financial intermediaries and are easily influenced by local governments, making them ideal partners for off-balance-sheet financing (Bellofatto & Besfamille, 2018). LCBs not only invest in government-issued bonds and trust products but also directly lend to LGFVs. As a result, LCBs' assets contain a significant amount of implicit local government debt, primarily in the form of long-term loans and investments, which exposes these banks to considerable government influence (Liu & Huang, 2023).

From a micro-level perspective, the central government's efforts to promote debt servicing through policies and supervision have not fundamentally improved local governments' fiscal conditions. Faced with political pressure to fulfill these tasks, local governments resort to new fundraising strategies, effectively shifting debt from explicit to implicit. As implicit debt grows, LCBs continue investing in government-issued financial products without conducting proper risk assessments, which weakens asset quality and heightens bank risk. Furthermore, since credit resources are finite, increased government-related lending inevitably crowds out private-sector credit (Yu et al., 2024). LCBs are the main providers of financial services to the local real economy, and credit crowding out exacerbates the financing difficulties of the real economy, which in turn adversely affects the development and debt-servicing capacity of the real economy and increases the risk of banks.

Finally, given the prominence of local government debt issues, existing literature has extensively explored its formation mechanisms and driving factors, such as regional competition (Qu et al., 2022), infrastructure development (Tsui, 2011), and political incentives (Li et al., 2024). Another body of research examines the negative consequences of local debt, including reduced corporate productivity (Zhu et al., 2022), environmental pollution (Zhou et al., 2023), and shadow banking risks (Chen et al., 2020). However, studies on the implications of debt servicing policies remain scarce, with only a few addressing long-term investment (Li et al., 2023) and employment (Ye et al., 2023). Therefore, detailed investigations into the impacts of debt servicing, especially from a banking perspective, warrant further research.

## 3. Research Methods and Design

### 3.1. Sample Selection and Data Source

The 2011 audit report by China's National Audit Office revealed that the debt balance in 2009 increased by 61.92% compared to 2008, marking the highest growth rate in recent years. In contrast, the debt balance in 2010 rose by 18.86% compared to 2009, the lowest growth rate in recent years. Given that local government debt financing in China has entered a phase of steady expansion since 2010, this study selects data from Chinese LCBs between 2010 and 2023 to evaluate and analyze the impact of government debt repayment on bank risk.

Local commercial banks include urban commercial banks and rural commercial banks, with their list and financial data sourced from the CSMAR database. To ensure the validity and reliability of the research data, the following data processing and cleaning procedures were applied: (1) Missing values in the explanatory and dependent variables were removed, and the dependent variables were winsorized at the 1% upper and lower levels. (2) Missing control variables for certain years were supplemented using interpolation and mean imputation methods. The final research sample comprises an unbalanced panel dataset containing 3,643 observations from 348 LCBs. According to statistics, our sample covers all LCBs of 206 cities across 29 provinces in China, providing substantial national representation.

### 3.2. Variable Setting

### 3.2.1 Dependent Variable

This study uses the non-performing loan ratio ($Npl_{i,j,t}$) as the dependent variable, a widely recognized indicator for assessing bank risk in the financial sector. Non-performing loans refer to loans that are overdue or unlikely to be fully repaid. The NPL ratio is calculated by dividing the balance of non-performing loans by the bank's total loan amount. In the robustness checks, the ratio of risk-weighted assets to total loans is used as an alternative indicator of bank risk.

### 3.2.2 Core Independent Variable

This study examines the impact of government debt servicing on the risk of LCBs, using the 2016 central government supervisory teams to seven provinces to oversee debt repayment as an exogenous shock. Consequently, the variable $Post_t$ is assigned a value of 1 for samples from 2016 onward and 0 otherwise. Similarly, the variable $Treat_j$ is set to 1 for banks located in the supervised regions and 0 otherwise. An interaction term between $Post_t$ and $Treat_j$ is constructed to identify the policy's effect. To address potential endogeneity, the robustness checks include an alternative control experiment targeting LCBs in the supervised regions.

### 3.2.3 Control Variables

Following previous studies (Altunbas et al., 2010; Khairi et al., 2021), several control variables are selected to enhance estimation accuracy. These variables capture aspects of profitability, asset quality, and bank fundamentals, specifically:

1. Bank asset size ($InAsset_{i,t}$): the natural logarithm of total assets.
2. Deposit-to-loan ratio ($DLratio_{i,t}$): the ratio of deposits to loans.
3. Capital adequacy ratio ($CaaR_{i,t}$): the ratio of net capital to risk-weighted assets.
4. Net interest income ($Inint_{i,t}$): the natural logarithm of net interest income.
5. Leverage ratio (Leverage): the ratio of total liabilities to total assets.

### 3.2.4 Descriptive Statistics

Table 1 presents the descriptive statistics. Analysis reveals that the average NPL ratio of LCBs from 2010 to 2023 is 2.092%, below the regulatory threshold of 5%, although the maximum value is approximately four times the mean, indicating high risk levels in some samples. The mean leverage ratio is 0.921, suggesting that most banks operate with high leverage. The significant variation in capital adequacy ratios also indicates good sample heterogeneity.

**Table 1. Descriptive Statistics**

| Variable | Obs | Mean | Std. dev. | Min | Max |
|---|---|---|---|---|---|
| **Npl** | 3642 | 2.092 | 1.644 | 0.29 | 8.73 |
| **Post*Treat** | 3642 | 0.156 | 0.363 | 0 | 1 |
| **InAsset** | 3642 | 24.821 | 1.391 | 19.276 | 28.952 |
| **DLratio** | 3642 | 1.534 | .542 | 0.134 | 13.472 |
| **CaaR** | 3642 | 14.031 | 3.09 | 1.98 | 59.61 |
| **lnint** | 3642 | 20.95 | 1.287 | 15.927 | 24.687 |
| **Leverage** | 3642 | 0.921 | 0.021 | 0.642 | 1.025 |

### 3.3 Empirical Model Setting

In order to verify the impact of government debt servicing on the risk of LCBs, this paper establishes a DID panel data model, as detailed in equation (1):

$$Npl_{i,j,t} = \beta_0 + \beta_1 Post_t * Treat_j + \beta_2 Control_{i,j,t} + \gamma_i + \delta_j + \varepsilon_t + \mu_{j,i,t} \quad (1)$$

In this equation, $Npl_{i,j,t}$ represents the delinquency rate of bank $i$ in place $j$ in year $t$, i.e., the explanatory variable. $Post_t * Treat_j$ is the explanatory variable, which is used to represent whether the policy is implemented or not. $Control_{i,j,t}$ is a set of control variables. $\gamma_i$, $\delta_j$ and $\varepsilon_t$ represent individual, regional and year level fixed effects. $\mu_{j,i,t}$ represents multidimensional random error term.

## 4. Results and Discussions

### 4.1 Benchmark regression results

The Column (1) of Table 2 represents the impact of government debt servicing policy on the risk associated with LCBs. Columns (2) to (4) display the regression results after the inclusion of control variables and fixed effects. The results indicate that the coefficients of the explanatory variables are all positive and significant at the 1% level, suggesting that government debt servicing increases the risk faced by LCBs.

For instance, in Column (4), the coefficient of the core explanatory variable is 0.453, indicating that, on average, the non-performing loan ratio of LCBs increases by 0.45 percentage points following the implementation of the policy. According to statistics, from 2010 to 2023, the average total loan amount for LCBs in China was 74.8 billion. Based on the regression results, each bank is projected to experience an increase of 340 million in non-performing loans annually.

**Table 2. Government debt servicing and bank risk**

| | (1) | (2) | (3) | (4) |
|---|---|---|---|---|
| Variables | Npl | Npl | Npl | Npl |
| Post*Treat | 0.459*** | 0.456*** | 0.859*** | 0.453*** |
| | (0.116) | (0.116) | (0.083) | (0.115) |
| lnAsset | | -0.0329 | -0.0139 | 0.309*** |
| | | (0.062) | (0.062) | (0.082) |
| DLratio | | -0.0948* | -0.155*** | -0.0908* |
| | | (0.049) | (0.052) | (0.047) |
| CaaR | | | -0.105*** | -0.0677*** |
| | | | (0.019) | (0.022) |
| lnint | | | -0.413*** | -0.423*** |
| | | | (0.065) | (0.088) |
| Leverage | | | -6.015*** | -10.47*** |
| | | | (2.221) | (2.805) |
| Constant | 2.041*** | 2.995* | 18.21*** | 13.93*** |
| | (0.025) | (1.535) | (2.080) | (2.740) |
| Fixed Effect | Yes | Yes | No | Yes |
| Observations | 3,684 | 3,670 | 3,727 | 3,643 |
| R-squared | 0.510 | 0.517 | 0.177 | 0.534 |

Notes: Robust standard errors in parentheses， *** p<0.01, ** p<0.05, * p<0.1

### 4.2 Endogenous concerns

The study by Demirci et al. (2019) indicates that local government debt in China exhibits endogeneity. Furthermore, other research has highlighted the presence of self-selection issues within Chinese policies (Wang & Yang, 2025). Although this paper incorporates a variety of control variables and accounts for three types of fixed effects in its regression analysis, concerns regarding endogeneity may still persist.

First, the State Council dispatched supervision teams to seven provinces and cities, including Beijing, Liaoning, Anhui, Shandong, Henan, Hubei, and Qinghai. As the figure 1 shows, these regions span China's central, western, and eastern areas, with significant differences in economic development and debt issues. For example, in 2016, Beijing's fiscal deficit ratio was approximately 5%, ranking at a relatively lowest level among the 31 provinces nationwide, while Qinghai's deficit ratio was around 50%, placing it the second highest in the country (National Bureau of Statistics of China, 2017). The wide variation in the geographical location and financial level of supervised areas explains to some extent the randomness of the pilot selection for the policy.

**Figure 1: 2016 Fiscal Deficit Rates for Supervised Areas and Nationwide**

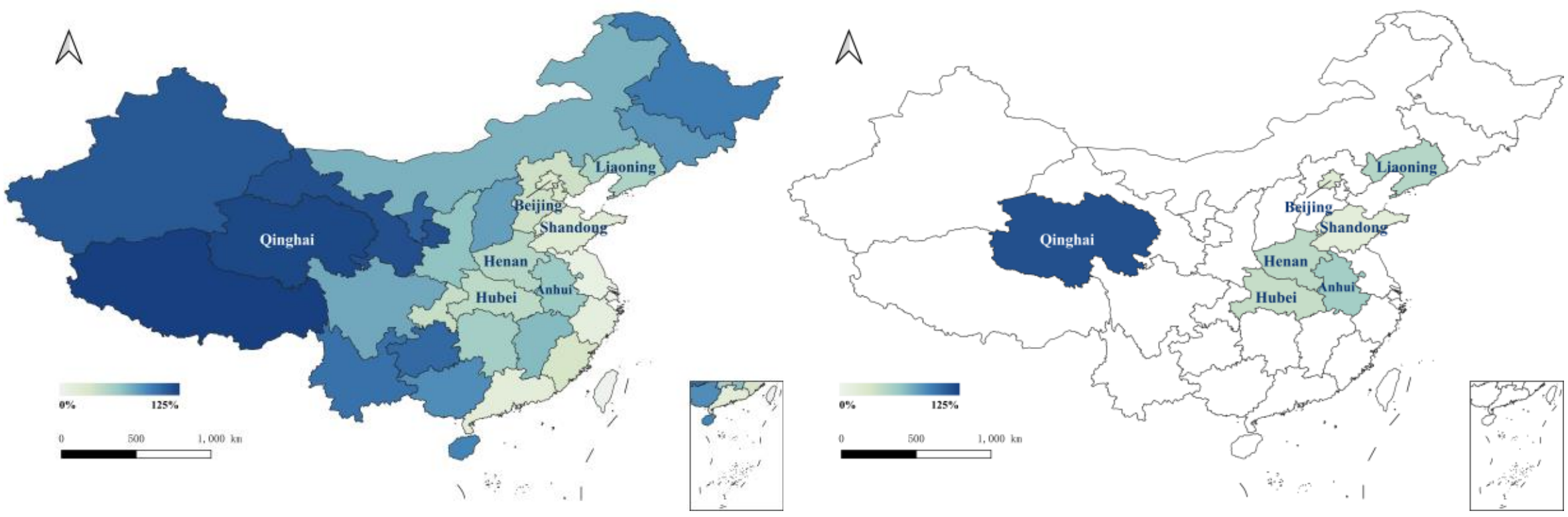

Then, this study employs both Propensity Score Matching (PSM) and a redesigned controlled experiment to address the endogeneity issues.

First, we conducted nearest neighbor matching at the bank level, with a ratio of 1:2. Table 3 presents the results of the balance test for Propensity Score Matching (PSM). The results indicate that prior to matching, the outcomes related to government debt servicing and bank risk were predominantly significant, suggesting substantial differences between the treatment and control groups. After matching, the t-test results for the covariates were all non-significant. Additionally, the post-matching bias remained within 10%. The Average Treatment Effect on the Treated (ATT) yielded a t-value of 7.17, which is significant at the 1% level, confirming the validity of the PSM results.

**Table 3. Balance Test for PSM**

| Variables | Matched/Unmatched | Mean | | %Bias | T-value |
|---|---|---|---|---|---|
| | | Treated | Control | | |
| lnAsset | Unmatched | 24.755 | 24.811 | -3.9 | -1.06 |
| | Matched | 24.754 | 24.854 | -7.000 | -1.530 |
| DLratio | Unmatched | 1.482 | 1.552 | -14.9 | -3.41*** |
| | Matched | 1.482 | 1.478 | 0.9 | 0.310 |
| CaaR | Unmatched | 13.69 | 14.129 | -13.9 | -3.76*** |
| | Matched | 13.693 | 13.671 | 0.7 | 0.160 |
| lnint | Unmatched | 20.767 | 20.986 | -16.2 | -4.51*** |
| | Matched | 20.770 | 20.854 | -6.200 | -1.340 |
| Leverage | Unmatched | 0.922 | 0.920 | 10.5 | 2.81*** |
| | Matched | 0.922 | 0.923 | -2.700 | -0.640 |

Notes: Robust standard errors in parentheses， *** p<0.01, ** p<0.05, * p<0.1

The Column (1) of Table 4 displays the results of the baseline regression, while Column (2) presents the regression results following PSM. These findings are consistent with those of the baseline regression and remain significant at the 1% level.

**Table 4. Regression results for endogeneity issues**

| | (1) Baseline | (2) PSM | (3) New DID |
|---|---|---|---|
| Variables | Npl | Npl | Npl |
| Post*Treat | 0.453*** | 0.544*** | 0.571** |
| | (0.115) | (0.139) | (0.245) |
| lnAsset | 0.309*** | 0.322** | 0.200 |
| | (0.082) | (0.126) | (0.128) |
| DLratio | -0.0908* | -0.482*** | -0.671*** |
| | (0.047) | (0.159) | (0.201) |
| CaaR | -0.0677*** | -0.0634* | -0.187*** |
| | (0.022) | (0.038) | (0.0253) |
| lnint | -0.423*** | -0.538*** | -0.145 |
| | (0.088) | (0.129) | (0.126) |
| Leverage | -10.47*** | -2.587 | -5.647 |
| | (2.805) | (4.849) | (3.628) |
| Constant | 13.93*** | 9.298** | 8.790** |
| | (2.740) | (4.624) | (3.510) |
| Fixed Effect | Yes | Yes | Yes |
| Observations | 3,643 | 1,985 | 880 |
| R-squared | 0.534 | 0.562 | 0.585 |

Notes: Robust standard errors in parentheses， *** p<0.01, ** p<0.05, * p<0.1

Secondly, considering the potential self-selection issues when the State Council dispatches supervisory teams (i.e., in policy pilot areas), it is possible that the local government debt issues in these supervised regions are inherently more severe, or that the LCBs exhibit stronger attributes of government financing tools. This could lead to biased estimates of the policy effects. Therefore, this study has designed a new controlled experiment.

Specifically, we selected samples from the supervised regions for regression analysis and redefined the $Treat_j$ variable. Given that multiple local government-controlled commercial banks operate in each city, their attributes related to financing tools and levels of government control differ. Accordingly, we categorized the LCBs in the supervised regions into two groups. Banks that operate beyond their province with numerous branches were designated as the control group. Although classified as LCBs, these institutions have broader business coverage and weaker ties to the government, resulting in a lower impact from government debt. For example, Bank of Beijing operates nearly 700 branches across more than ten provinces and is a publicly listed bank, thereby experiencing minimal influence from the Beijing government. Conversely, LCBs with fewer branches and limited operational scope within their city or province were classified as the treatment group, which is due to their close relationship with local government debt and stronger susceptibility to policy impacts. Ultimately, seven banks from pilot province are selected as the control group, while the others form the treatment group. After generating a new interaction term with the $Post_t$ variable, we conducted the regression.

The Column (3) of Table 4 presents the regression results of the new controlled experiment, which are consistent with the baseline regression and remain significant at the 5% level. The regression results indicate that the local government debt servicing policy indeed expands the risk faced by local commercial banks, and the endogeneity issue has been addressed.

### 4.3 Robustness Tests

#### 4.3.1 Parallel Test

In order to investigate the dynamics of bank NPLs after the policy was implemented and to verify that the increase in bank NPLs was not due to pre-existing ex-ante trends, this paper builds on previous research (Kudamatsu, 2012) by applying the event study methodology to conduct a parallel test, and develops the following model:

$$Npl_{i,j,t} = \alpha + \sum_{\substack{k=-6 \\ k\neq 0}}^{6} \vartheta_k D_{j,t}^k + \theta Control_{i,j,t} + \gamma_i + \delta_j + \varepsilon_t + \mu_{j,i,t} \tag{2}$$

In equation (2), considering that the policy was only implemented at the end of 2016, the impact on the current year of 2016 is relatively small, so we set 2016 as the base year for $s$. When $t - s = k$, $D_{j,t}^{k} = 1$, otherwise, $D_{j,t}^{k} = 0$. In addition, in order to have a same distribution of periods before and after the policy, and also to have a relatively consistent sample size across periods. When $t - s > 6$, let $D_{j,t}^{6} = 1$.

**Figure 2: Parallel Test**

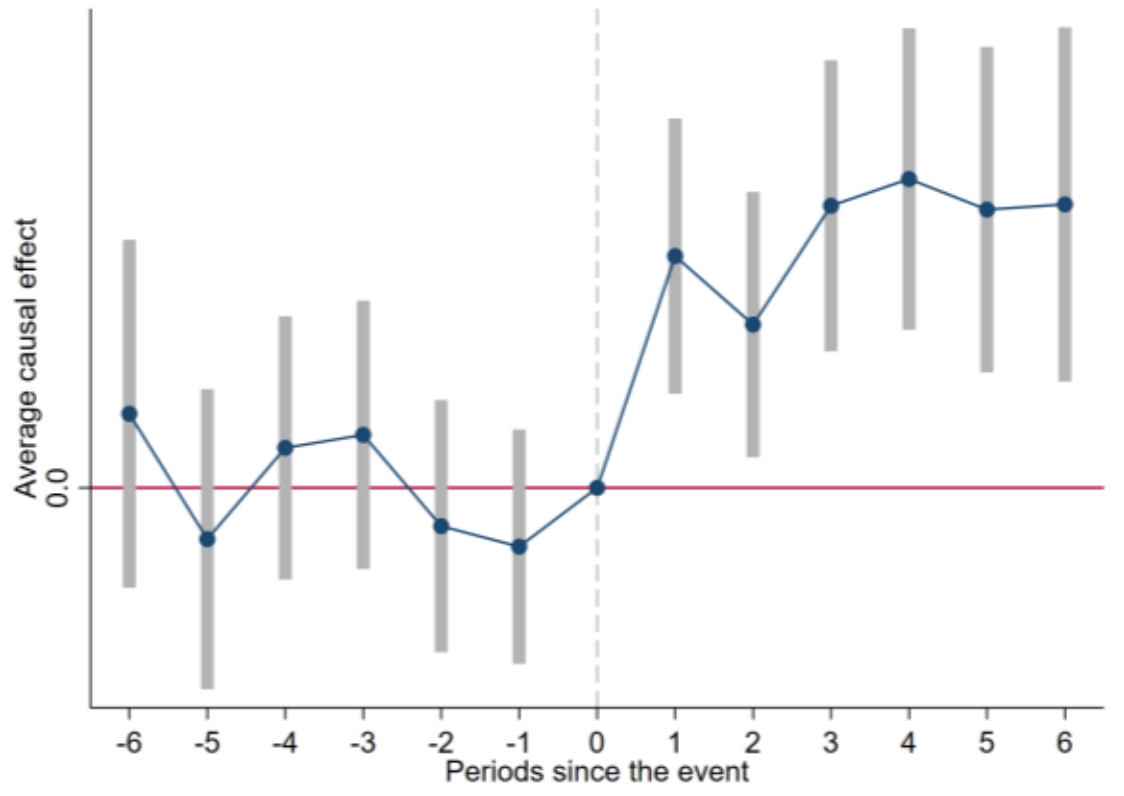

**Figure 3: Placebo test**

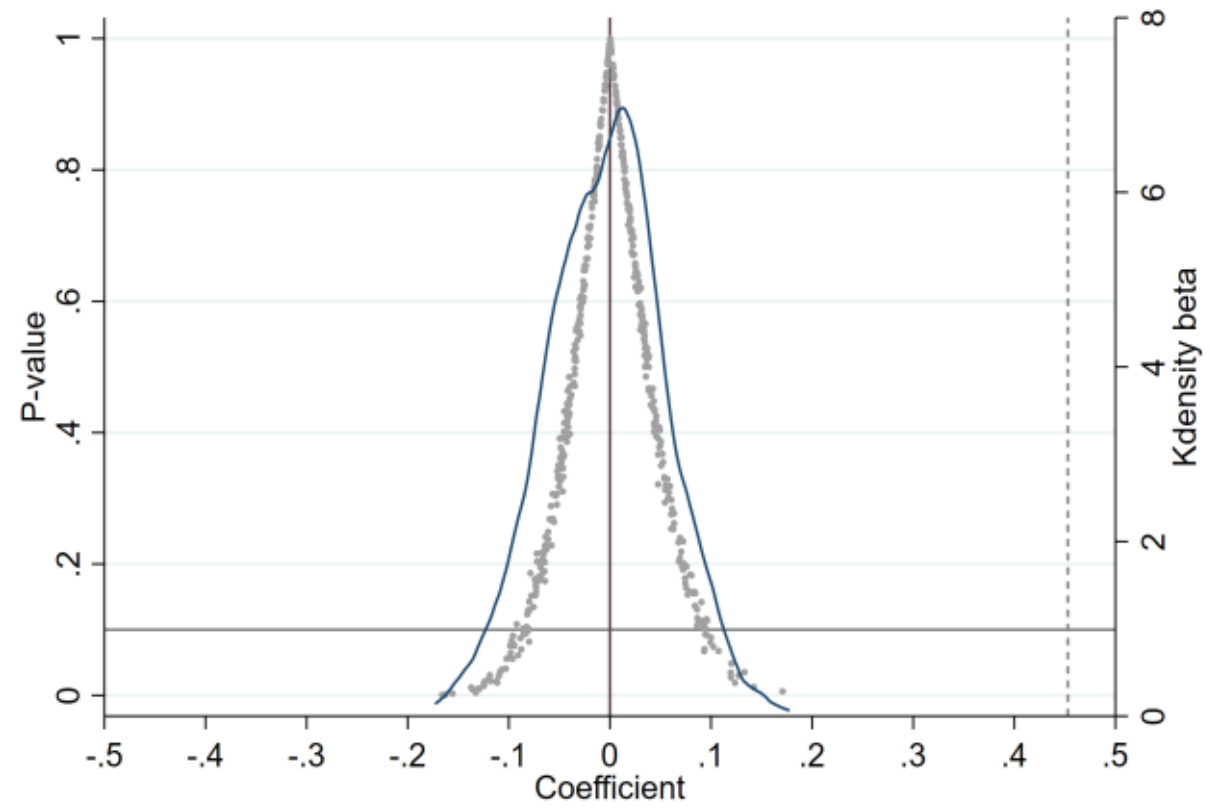

The results presented in Figure 2 indicate that prior to the baseline period, the regression results were not significant, suggesting that there were no notable differences between the treatment and control groups before the policy implementation. In contrast, after the baseline period, the regression results became significant, with positive coefficients, indicating that the policy has increased bank risk in the implementation areas, which is consistent with the findings of the baseline regression.

#### 4.3.2 Placebo test

To further verify that the increase in the NPL ratio is attributable to the local government debt servicing policy rather than other contemporaneous observational factors, this study randomly assigned policy occurrence years between 2010 and 2023. Additionally, we randomly selected 7 out of the 31 provincial administrative regions in China as implementation areas, repeating this process 500 times for a placebo test. The results of the test are illustrated in Figure 3.

The findings in Figure 3 indicate that the mean of the coefficient density distribution and the P-value distribution from the 500 simulated regressions is close to zero, and they generally fit a normal distribution. The coefficient from our baseline regression is 0.453, positioned on the right side of the simulated regression distribution. Therefore, the results of the baseline regression are robust.

### 4.4 Other robustness tests

In addition to placebo tests and parallel test, this study makes targeted adjustments to the sample and variables to mitigate potential biases in model estimation due to sample and data issues. The specific details are as follows:

First, while the baseline regression uses the NPL ratio to measure bank risk levels, drawing on previous research (Ferri & Pesic, 2017), this study adopts the proportion of risk-weighted assets ($Risk_{i,j,t}$) as a substitute variable. Risk-weighted assets represent the weighted sum of various asset risk coefficients, which are widely used to reflect a bank's risk level. The results are presented in the Column (1) of Table 5.

Second, considering the significant economic downturn during the pandemic, which adversely affected corporate operations and government finances, leading to a higher incidence of loan defaults and increased NPL ratios, this study excludes the years during the pandemic. Regression analysis is thus conducted using data solely from 2010-2019 and 2023. The regression results are shown in the Column (2) of Table 5.

Furthermore, there exist substantial developmental disparities among the 31 provincial administrative regions in mainland China. The western regions and Guizhou Province exhibit lower levels of economic development, lower per capita GDP, larger local government debt scales, higher deficits, and greater repayment pressures. Theoretically, NPLs in these regions are more closely related to local government debt, and the policies aimed at alleviating local government debt will have a more pronounced impact on the NPL ratios of commercial banks in these areas. Consequently, this study excludes local commercial banks located in the western regions and Guizhou Province from the sample for regression analysis. The results are displayed in the Column (3) of Table 5.

The results presented in Table 5 are consistent with the baseline regression and remain significant at the 1% level, indicating that the findings of the baseline regression are robust.

**Table 5. Regression Results of the Robustness Test**

| Variables | (1) New Variable<br>Risk | (2) Non-Covid<br>Npl | (3) Non-Western<br>Npl |
|---|---|---|---|
| Post*Treat | 0.090*** | 0.315*** | 0.483*** |
| | (0.028) | (0.117) | (0.125) |
| lnAsset | 0.0366 | 0.339*** | 0.263*** |
| | (0.038) | (0.084) | (0.0968) |
| DLratio | 0.0785** | -0.118* | -0.229** |
| | (0.032) | (0.068) | (0.109) |
| CaaR | -0.0113*** | -0.0662*** | -0.0780*** |
| | (0.003) | (0.023) | (0.028) |
| lnint | -0.0448 | -0.416*** | -0.495*** |
| | (0.037) | (0.100) | (0.104) |
| Leverage | -2.687*** | -11.50*** | -9.619*** |
| | (0.635) | (3.066) | (3.578) |
| Constant | 2.726*** | 14.06*** | 16.14*** |
| | (0.849) | (2.950) | (3.408) |
| Fixed Effect | Yes | Yes | Yes |
| Observations | 3,477 | 2,800 | 3,056 |
| R-squared | 0.550 | 0.542 | 0.525 |

Notes: Robust standard errors in parentheses, *** p<0.01, ** p<0.05, * p<0.1

### 4.5 Heterogeneity Analysis

In the preceding sections, the study examined how local government debt servicing affects the risk of local banks using the entire sample. A series of methods were employed to rigorously test the robustness of the research findings, confirming the correlation between them. Furthermore, this section utilizes grouped regression methods to investigate the heterogeneous effects of local government debt servicing on bank risk from three perspectives: customer concentration, bank type, and coverage scale. This approach captures varying degrees of association between different types of banks and local governments, allowing for a more in-depth analysis of the differential impacts of local government debt servicing on local bank risk. Finally, interaction terms are used to verify whether the inter-group differences are significant. The results of the heterogeneity analysis are presented in Table 6.

First, loan concentration risk is one of the primary causes of banking crises. Existing research indicates that banks can reduce credit default risk by lowering customer concentration (Diamond, 1984). High loan concentration means that commercial banks are significantly impacted by the performance of a single client. During economic downturns and periods of market volatility, excessive concentration of credit can render banks less flexible in addressing government debt issues. The results presented in the Column (1) of Table 6 indicate that banks with higher customer concentration are more significantly affected by government debt.

Second, local commercial banks can be divided into urban commercial banks and rural commercial banks. Urban commercial banks tend to be larger, with more branches and a broader business scope, alongside a more diversified ownership structure, resulting in relatively less influence from local governments. In contrast, rural commercial banks are more susceptible to government control and often serve as financing tools. The results in the Column (2) of Table 6 show that the risk of rural commercial banks is more significantly impacted by local government debt servicing.

Finally, the number of branches represents the geographic coverage of a bank's operations. Banks with fewer branches typically focus on local operations and are thus more strongly influenced by local governments. The grouped regression results in the Column (3) of Table indicate that government debt servicing has a greater impact on the risk of banks with fewer branches.

**Table 6. Heterogeneity Analysis**

| | (1) Concentration | | (2) Type | | (3) Branch | |
|---|---|---|---|---|---|---|
| Variables | >20% | <20% | City | Rural | >68 | <68 |
| Post*Treat | 0.504*** | 0.783* | -0.0445 | 1.148*** | 0.246 | 0.665*** |
| | (0.138) | (0.418) | (0.151) | (0.210) | (0.153) | (0.234) |
| lnAsset | 0.374*** | 0.081 | 0.563*** | -0.012 | 0.0497 | 0.505*** |
| | (0.115) | (0.167) | (0.175) | (0.105) | (0.135) | (0.129) |
| DLratio | -0.091 | -0.051 | -0.057 | -0.014 | -0.0901 | -0.116 |
| | (0.067) | (0.071) | (0.065) | (0.049) | (0.0633) | (0.0733) |
| CaaR | -0.003 | -0.074** | 0.009 | -0.043* | -0.010*** | -0.0771** |
| | (0.029) | (0.033) | (0.042) | (0.023) | (0.026) | (0.0300) |
| lnint | -0.575*** | 0.149 | -0.732*** | -0.130 | -0.343*** | -0.639*** |
| | (0.127) | (0.183) | (0.169) | (0.109) | (0.115) | (0.163) |
| Leverage | -5.432 | -5.108 | -4.808 | -0.467 | -1.250 | -16.38*** |
| | (3.563) | (4.936) | (5.103) | (3.565) | (3.767) | (3.984) |
| P-Interaction | 0.013** | | 0.000*** | | 0.096* | |
| Constant | 9.920*** | 2.275 | 7.439 | 6.135* | 10.60*** | 19.37*** |
| | (3.635) | (3.861) | (5.422) | (3.218) | (3.909) | (3.987) |
| Fixed Effect | Yes | Yes | Yes | Yes | Yes | Yes |
| Observations | 2,192 | 687 | 1,348 | 1,574 | 1,837 | 1,749 |
| R-squared | 0.523 | 0.457 | 0.312 | 0.565 | 0.565 | 0.586 |

Notes: Robust standard errors in parentheses， *** p<0.01, ** p<0.05, * p<0.1

## 5. Further Mechanism Analysis

The theoretical analysis section explains how local government debt servicing can produce counterproductive effects that increase bank risk. To further verify the economic mechanisms involved, this section conducts empirical analysis using data. The impact mechanisms are presented from two perspectives: the role of banks as financing tools and the crowding-out effect of government on private credit.

### 5.1 Financing tool

As previously mentioned, LCBs primarily provide financing support to the government by purchasing financial products such as government-issued bonds and trusts. In Shandong Province, one of the regions under supervision, over 90 billion yuan in debt was serviced in the first year of policy implementation, with approximately 58.8% achieved through the issuance of replacement bonds (Shandong Province Government, 2017). This indicates that the funding required for debt servicing may necessitate the issuance of bonds, trusts, and other financial products by the government, with LBCs being the main investors in these products.

Based on this, this study constructs a new variable $Goinv_{i,j,t}$, by calculating the sum of banks' receivable investments (such as trusts) and debt investments (such as bonds) to measure the role of banks as government financing tools. The regression results are presented in the Column (1) of Table 7.

The regression results indicate that after the policy implementation, the role of local commercial banks as government financing tools has strengthened, and this effect is significant at the 1% level, consistent with our theoretical analysis.

**Table 7. Mechanism Analysis**

| | (1) | (2) |
|---|---|---|
| Variables | Goinv | Coloan |
| Post*Treat | 16.30*** | -0.010** |
| | (5.451) | (0.004) |
| lnAsset | 21.20*** | -0.005 |
| | (2.170) | (0.004) |
| DLratio | 0.610 | 0.005** |
| | (1.030) | (0.002) |
| CaaR | -0.149 | 0.000 |
| | (0.336) | (0.001) |
| lnint | 7.208*** | 0.010*** |
| | (1.965) | (0.003) |
| Leverage | -395.4*** | 0.140 |
| | (78.26) | (0.133) |
| Constant | -289.0*** | 0.090 |
| | (69.42) | (0.106) |
| Fixed Effect | Yes | Yes |
| Observations | 2,898 | 2,316 |
| R-squared | 0.664 | 0.929 |

Notes: Robust standard errors in parentheses， *** p<0.01, ** p<0.05, * p<0.1

### 5.2 Credit Crowding Out

In addition to directly investing in government-issued financial products, local commercial banks also provide loans directly to LGFVs to offer financing support. In a context of limited credit resources, loans to LGFVs can crowd out local private credit or increase financing costs (Huang et al., 2020; Zhang et al., 2022), thereby impacting the operational capacity and debt repayment ability of the real economy, which in turn increases bank risk.

To this end, we use the ratio of consumer loans to total loans to represent private bank credit, with the regression results presented in the Column (2) of Table 7. The regression results indicate that after the policy implementation, the proportion of banks' consumer loans has decreased and is significant at the 5% level, demonstrating the existence of credit crowding out, which aligns with our analytical findings.

## 6. Conclusion

In the context of local government debt servicing tasks in China, the government obtains funding through off-balance-sheet financing to repay on-balance-sheet explicit debts. In this process, local governments successfully transfer fiscal risks to the financial sector. The risk level of local commercial banks, as the main platform for assuming government debt, has also increased. We explore the impact of government debt servicing policies on local bank risk by constructing a simple difference-in-differences model, utilizing data from 348 local commercial banks in China from 2010 to 2023 for empirical testing.

The research findings indicate that local government debt servicing policies significantly increase the risk faced by local commercial banks. We conducted a series of robustness checks, including addressing endogeneity issues, conducting parallel trend tests, placebo tests, replacing the dependent variable, and altering the sample and time ranges, and the results remain robust. Heterogeneity analysis reveals that the increase in risk is more pronounced among rural commercial banks, banks with high customer concentration, and those with fewer branches. Further mechanism analysis indicates that local debt servicing primarily increases bank risk by compelling banks to act as financing tools, thereby deteriorating asset quality and crowding out private credit, which reduces local repayment capacity.

This study investigates the impact of government debt servicing on bank risk, holding significant practical and policy implications for government governance, bank operations, and debt servicing. The central government should consider formulating more comprehensive and systematic debt servicing policies. Furthermore, there is a need to enhance the transparency of debt and diversify its sources, along with strengthening debt management. Finally, the financing tool attributes of local commercial banks should be controlled to promote their role in the development of the local real economy.

However, this study also has its limitations. Future research should consider constructing theoretical models to further enhance the completeness of the study and attempt to fit empirical data to these theoretical models to strengthen the robustness of the conclusions.